\documentclass[12pt,a4paper]{article}

\usepackage{fullpage,amsmath,amssymb,amsfonts,epsfig,setspace}

\psfigdriver{dvips}

\newcommand{\beq}{\begin{equation}}
\newcommand{\eeq}{\end{equation}}
\newcommand{\la}{\langle}
\newcommand{\ra}{\rangle}

\newcommand{\bea}{\begin{eqnarray}}
\newcommand{\eea}{\end{eqnarray}}

\usepackage{graphicx}
\usepackage{subfigure}
\usepackage{amsmath}

\begin{document}

\title{\begin{bfseries}{Emergence of Network Structure in Models of Collective Evolution and Evolutionary  Dynamics.}
\end{bfseries}}

\author{Henrik Jeldtoft Jensen\footnotemark[1] \footnotemark[2] \footnotemark[3] }

\maketitle

\fnsymbol{footnote}
\footnotetext[1]{Institute for Mathematical Sciences, Imperial College London, 53 Prince's Gate, South Kensington campus, SW7 2PG, UK.}
\footnotetext[2]{Department of Mathematics, Imperial College London,  South Kensington campus,  London SW7 2AZ, UK.}
\footnotetext[3]{Email and URL (h.jensen@imperial.ac.uk; www.ma.ic.ac.uk/$^\sim$hjjens)}

\begin{abstract}
We consider an evolving network of a fixed number of nodes. The allocation of edges is a dynamical stochastic process inspired by biological reproduction dynamics, namely by deleting and duplicating existing nodes and their edges. The properties of the degree distribution in the stationary state is analysed by use of the Fokker-Planck equation. For a broad range of parameters exponential degree distributions are observed. The mechanism responsible for this behaviour is illuminated by use of a simple mean field equation and reproduced by the Fokker-Planck equation treating the degree-degree correlations approximately. In the limit of zero mutations the degree distribution becomes a power law.
\end{abstract}
\noindent {\bf Keywords:} Networks, Dynamics, Evolution, Degree distribution\\ 
\section{Introduction: Networks and evolutionary dynamics}
Whenever a phenomena can be thought of in terms of components and relations between components the mathematical language of graph theory or networks may be helpful to the description, analysis and the understanding of the relevant problem of interest. A large amount of work is currently being done with the aim to understand the structure and statistical properties of networks in the hope that certain aspects of the general mathematical characterisation of network structure may be related to common functional properties, e.g. vulnerability to breakdown of part of the network \cite{Biggs,Albert_Barabasi,Newman_rev}. 

Our aim in the present paper is to discuss an example of a network, inspired by evolutionary dynamics where we can relate the emergent network structure to microscopic details of the reproductive dynamics generating the network. We think of the network as arising from the dynamics of reproducing individuals linked together with other individuals in ways that depend on their type. The nodes represents the species emerging in some type space as a result of the mutation prone reproductive dynamics at the level of individuals. As individuals of various types interact and reproduce they segregate in type space allowing local regions of high occupancy to be identified as species. In this way a network structure emerges on the space of types. Nodes consists of occupied, or extant, species, and edges connecting the these nodes represent the interactions between the species. The basic dynamics is individually driven and consists of individuals being born for then to disappear again somewhat later at random killing event (driven as a Poisson process). At the level of individuals the dynamics doesn't appear very exciting. However at the systems level, the dynamics of the occupancy density in type space exhibit interesting emerging structures. We will focus on the functional form of the degree distribution and its relation to the amount of mutation involved in the underlying reproductive dynamics. 

Stumpf and co-workers have studied the properties of sub-networks obtained by random sampling nodes in a larger network \cite{Stumpf:PRE_05}. They showed that only binomial (or Poisson) networks are invariant under decimation. If the large network has a binomial degree distribution a randomly sampled subset of the network will also exhibit a binomial degree distribution. Here we are concerned with evolutionary dynamics on very large and correlated networks. The dynamics generates sub-networks typically with exponential degree distributions even when the full network has a binomial degree distribution.   

The paper is organised as follows. To set the scene we first briefly describe the individual based Tangled Nature model of evolutionary ecology and review a few aspects of the phenomenology of the networks emerging at the level of types \cite{TaNaNetwork,lair05:tang,lair07:EcoMod}. Next we use a simplified model defined at the level of types - or species - to gain some understanding of how evolutionary dynamics on correlated networks can give rise to exponential networks and, in certain limits, to a power law degree distribution.

\section{Tangled Nature model}\label{TaNa_model} 
The basic of the Tangled Nature model \cite{chri02:tang,lair05:tang,lair07:EcoMod} is as follows. Individuals, $\{\alpha, \beta, ...\}$ are described by type vectors $\bf{T}^{\alpha}=(\it{T}^{\alpha}_{1},\it{T}^{\alpha}_{2},...,\it{T}^{\alpha}_{L})$. The number of individuals of type ${\bf T}$ at time $t$ is denoted by $n({\bf T},t$. Different types influence each other through an interaction matrix ({\bf J}-matrix) that accounts for all possible interactions between any possible set of types. Selection leads to only a small fraction of types being occupied and their interactions will be described by a small subset of the elements of this complete matrix. The structure of the interaction network between extant types is found to depend on the statistical properties of {\bf J}. A proportion, $\theta$, of the elements of the {\bf J}-matrix, ${\rm{\bf J}}({\rm{\bf T}}^{\alpha}, {\rm{\bf T}}^{\beta})$ are assigned non-zero (an non symmetric) values all other elements  are zero. The interactions assigned in the type space can either be uncorrelated \cite{hall02:time,TaNaNetwork}, or correlated \cite{lair05:tang,lair07:EcoMod}. If no correlations are present in the type space the evolved networks of interactions between extant species exhibit a binomial degree distribution as does the underlying network of non-zero {\bf J}-matrix elements \cite{TaNaNetwork}. The correlated case is more interesting and the one we will concentrate on in this paper. Correlations are made to decay exponentially with separation in type space. This implies that off-spring will see a set of interactions which are very similar to the interactions of the parent even when mutations make the off-spring differ from the parent. When correlations are present in {\bf J} the evolutionary dynamics generates networks with exponential degree distributions in contrast to the binomial distributions exhibited by a network constructed by randomly seleceting positions in type space. 

The dynamics of the system consists of the following time step. First randomly choose an individual and attempt to remove the individual with constant probability $p_{kill}$ that is independent of type and time. Next randomly choose an individual and replicated it with a probability that depends on the type of the individual and the configuration in type space. In the Tangled Nature model a probabilistic weight is calculated that depends on how the type of the chosen individual interact with other extant types. This value is determined through the use of a weight function,
\begin{equation}
\it{H} ({\bf T}^{\alpha},\it{t}) = 
  a_{1} \frac{\sum_{{\bf T}\in {\cal T}} {\bf J} ({\bf T}^{\alpha}, {\bf T}) n({\bf T},t)} {\sum_{{\bf T}\in {\cal T}} {\bf C} ({\bf T}^{\alpha}, {\bf T}) n({\bf T},t)} 
- a_{2} \sum_{{\bf T}\in {\cal T}} {\bf C} ({\bf T}^{\alpha}, {\bf T}) n({\bf T},t) 
- a_{3} {\frac{N(t)}{R(t)}}.\label{eq.hamiltonian3}
\end{equation}
This is monotonically mapped to the interval [0,1], appropriate for a probability measure, by using the following function,
\begin{equation}
P_{repro}=\frac{\exp[{\it H}({\bf T}^{\alpha},{\it t})]}{1+exp[{\it H}({\bf T}^{\alpha},{\it t})]}.\label{eq.compfermi}
\end{equation}

The sums of Eq.(\ref{eq.hamiltonian3}) are made over the points in type space, ${\cal T}$, and the occupancies, $n({\bf T},t)$ are used to account for the multiplicity of individuals with the same type vector. We imagine a well mixed system and otherwise neglect spatial aspects. We stress that the type space is a pre-defined, complete set of all {\em possible} types and it is evolution and contingency that select the actualised types in the evolved system. The {\bf{J}}-matrix is similarly a pre-defined complete set of all possible interactions for all possible types that may exist {\emph{in potentia}}. The elements of the matrix are exponentially correlated in the following way \cite{lair05:tang}
\begin{equation}
{\bf C} ({\bf J}({\bf T }^{\alpha},{\bf T }^{\gamma}),{\bf J }({\bf T }^{\beta},{\bf T}^{\gamma})) = \exp[-{\Delta}({\bf T}^{\alpha},{\bf T}^{\beta}) /\xi]\,\,{\in[0,1)},
\label{eq.correlation} 
\end{equation}
Here, $\xi$ is the correlation length and ${\bf{C}} ({\bf{J}}({\bf{T}}^{\alpha},{\bf{T}}^{\gamma}),{\bf{J}}({\bf{T}}^{\beta},{\bf{T}}^{\gamma}))$ is the correlation between the interaction strengths of two types ${\bf T}\alpha$ and ${\bf T}\beta$ when each are interacting with a third, ${\bf T}\gamma$.

The evolutionary dynamics generates a set of occupied types linked together according to the interaction matrix ${\bf J}$. We are here concerned with the properties of the degree distribution of this interaction network. Simulations found that the degree distribution $P(k)$ is of exponential form $P(k)\propto \exp(-k/k_0)$ whenever the interaction matrix is correlated \cite{lair05:tang}. 

\section{Simple node model}\label{simple_model}
To understand this phenomenology we now neglect the fluctuations present at the level of individual based dynamics and assume a more coarse grained view point in which we consider types as either occupied or not. I.e. we turn the measure $n({\bf T},t)$ into a binary equal to 1 when $n({\bf T},t)>0$ and zero when $n({\bf T},t)=0$. We consider the dynamics at the level of types which implies that creation events correspond to one type splitting into two types (a speciation event) and annihilation events correspond to a type going extinct. We elevate the dynamics of the individual based Tangled Nature model to the level of types in the following way. 

For simplicity we consider a fixed number $D$ (for diversity) of types. A timestep consists of choosing a type, or node, at random and remove it together with all its associated edges. Next another node, a {\emph{parent}} type, is randomly selected from the remaining $D{-}1$ nodes and is duplicated in the form of a {\emph{daughter}}. All types connected to the parent are now given connections to the daughter with probability $P_{e}$. All types unconnected to the parent are given connections to the daughter with probability, $P_{n}$. An edge between the daughter and parent is placed with probability $P_{p}$\footnote{For a similar model with $P_n=0$ see \cite{Farid:NJP06}}. These probabilities represent the similarity or correlations between offspring and parent induced in the Tangled Nature model by the correlations in the interaction matrix ${\bf J}$. 

\begin{figure}[h]
\centering
\includegraphics[width=8cm]{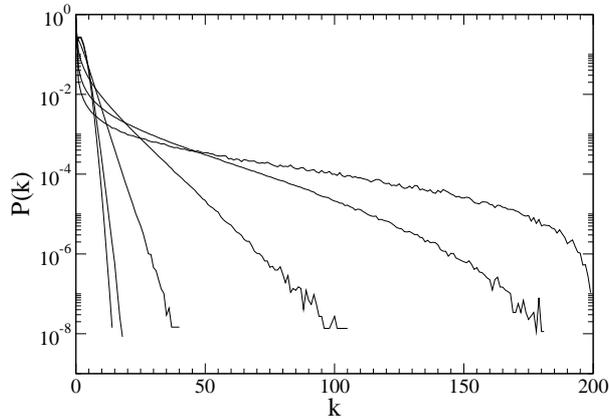}
\caption{Simulated degree distributions of the node model for $D=200$ and $P_p=0.01$ using the imperfect duplication process. From short to long tail we have $P_{e}=$ 0.01, 0.25, 0.75, 0.95, 0.99, 0.999 and $P_n$ was chosen to be $P_n=P_p(1-P_e)/(1-P_p)$ in order to keep the connectance fixed (Laird \& Jensen, 2007).} 
\label{simulations}
\end{figure}

This procedure is simple to simulate and produces, independent of initial configuration, after a relatively short transient, networks with degree distributions that behave exponentially to a very good approximation except in the limit $P_e\rightarrow 1$ and vanishing $P_n$ and $P_p$. We will in a moment write down the complete Fokker-Planck equation for the degree distribution of a network evolving according to this process. The full equation is, however, rather involved and can only be solved by numerical iteration. It is therefore illuminating to make the following simplistic and heuristic considerations. Let $n_k(t)$ denote the number of nodes of degree $k$ after $t$ timesteps. Let us focus solely on the following three aspects of the dynamics: 1) a node of degree $k$ is selected for annihilation, this occurs with probability $n_k(t)/D$. 2) a parent node of degree $k$ receives an edge to the daughter, this occurs with probability $P_pn_k(t)/D$. 3) A node receives an edge because it is selected to become a neighbour of the daughter (of a degree $k_p$ parent) with probability 
\beq
(P_e+P_n)[k_p/D][n_{k_p}(t)/D]\mapsto (P_e+P_n)[\la k\ra/D].
\eeq 
As we are seeking a qualitative mean field equation, we substituted the average degree $\la k\ra$ in the last expression. Note we have deliberately neglected the direct effect of the daughter. The rationale behind this is that there is only one daughter node, but the daughter has typically several neighbour nodes. We combine these events to obtain the following iterative equation.
\beq
\label{mean_field_Fokker_Planck}
n_k(t+1)=n_k(t) - \frac{n_k(t)}{D} +\frac{P_p+(P_e+P_n)\la k\ra}{D}[n{k-1}(t)-n_k(t)]. 
\eeq 
The stationary solution to this equation is readily obtained self-consistently to be 
\beq
n(k)=n(0)\exp[-k/k_0],
\eeq
with
\beq
k_0=1/\ln(1+[P_p+(P_e+P_n)\la k\ra]^{-1}),
\eeq
and 
\beq
\la k\ra=D(\frac{D}{n(0)}-1).
\eeq
The dependence on $P_e$ of $k_0$ in this solution is in qualitative agreement with the change in the exponential part of the degree distribution obtained in simulations of the network, see Fig. \ref{simulations}. However, this simplistic mean field discussion is only of heuristic value. We now present the full Fokker-Planck equation for the process.
\vspace{1cm}

\section{Fokker-Planck equation}
Let us now develop the Fokker-Planck equation for the ensemble averaged time dependent number of nodes of degree $k$, $n_{k}(t)$, constrained by the condition, $\sum_{k}n_{k}(t)=D$. We structure the analysis in the following way. {\em Removal} (R): The effect of removing,  from a population of $D$, a node and its edges described by a rate term $\Gamma_R(D,k,t)$. {\em Duplication} (D): The effect of introducing, into a population of $D-1$, a new daughter node and attaching edges described by a rate term $\Gamma_{Du}(D-1,k,t)$. Our equation has accordingly the form.
\begin{equation}
n_{k}(t+1)=n_{k}(t) + \Gamma_R(D,k,t) + \Gamma_{Du}(D-1,k,t). 
\label{master_D}
\end{equation}
Removal of a node affects the network in two ways: ($\Gamma^r_R$) the node removed from the network and ($\Gamma^a_R$) the effect on the nodes adjacent, i.e. sharing edges, with the node being removed. Therefore
\beq
\Gamma_R(k)= -\Gamma^r_R(k) +\Gamma^a_R(k{+}1)-\Gamma^a_R(k).
\eeq
 The effect of the duplication process is conveniently broken up into three sub-effects: ($\Gamma^p_{Du}$) the effect on the parent, ($\Gamma^d_{Du}$) the effect on the daughter, ($\Gamma^a_{Du}$) the effect on the adjacent nodes, i.e. those that will receive an extra edge as a result of the duplication. Hence
\beq
\Gamma_{Du}(k) = \Gamma^p_{Du}(k{-}1)-\Gamma^p_{Du}(k)+\Gamma^d_{Du}(k)+\Gamma^a_{Du}(k{-}1)-\Gamma^a_{Du}(k).    
\eeq
We have suppressed the timestep, $t$, and network size, $D$, for notational ease. 

Next we derive detailed expressions for each of the terms above.
The direct effect on $n_{k}$ of removing a node of degree, $k$ is to decrement its value. The probability of selecting a node of degree $k$ is ${n_{k}/N}$, and therefore,
\begin{equation}
\Gamma^r_R(k)=\frac{n_{k}}{N}.
\end{equation}
After the removal, the degree of the nodes connected to the removed node, i.e. the adjacent nodes, will decrease by one. For this we need the {\em Edge} probability, 
\beq
\label{P_Ed}
P_{Ed}(k_1,k_2,q)={\rm Prob}\{{\rm node\; of\; degree\;} k_{1}\; {\rm is\; connected\; to\;} q {\rm \;  nodes\; of\; degree\;} k_{2}\}. 
\eeq
In general we do not have an analytic expression for $P_{Ed}(k_1,k_2,q)$, but below we give approximate forms neglecting, or treating non-rigousrly, degree-degree correlations. Here we note 
\begin{equation}
\Gamma_R^a(k)=\sum_{k_r=1}^{N{-}1}\frac{n_{k_r}}{N}\sum_{q=1}^{k_r}qP_{Ed}(k_r,k,q).
\end{equation}
The first sum is over the degree of the removed node, the second sum is over the number, $q$, of nodes of degree $k=0,1,...,N{-}1$ the removed node is connected to.

A node of degree $k$ is selected for duplication with probability $n_k/(D-1)$. The daughter of this node receives an edge to the parent with probability $P_p$. Thus the parent increases its degree by one with probability
\begin{equation}
\Gamma_{Du}^p(k)=P_p\frac{n_k}{D-1}.
\end{equation}
The new daughter node can add to $n_k$ by an amount determined by the probability of finishing with $k$ edges,
\begin{equation}
\Gamma_{Du}^d(k) =P_p\Lambda(k-1)+(1-P_p)\Lambda(k).
\end{equation}
To keep track of the bookkeeping we have introduced a new probability
\begin{equation}
\Lambda(k) =
{\rm Prob}\{{\rm daughter \; receives \; {\it k} \; edges \; to \; nodes  \; different \; from \; the \; parent} \},
\end{equation}
which is given by 
\begin{equation}
\Lambda(k)= \sum_{k_p=0}^{D-2}\sum_{k_1=0}^{{\rm min}[k_p,k]}\sum_{k_2=0}^{{\rm min}[D-2-k_p,k]}\frac{n_{k_p}}{D-1}\delta(k_1{+}k_2{-}k)\Omega(k_1,k_2,k_p),
\end{equation}
The right hand side adds up the probabilities associated with the process where the daughter inherits $k_1$ edges to nodes already connected to the parent. Each of these edges is inherited by the daughter with probability, $P_{e}$. The daughter may receive an additional $k_2=k-k_1$ edges to nodes not connected to the parent. Each of these edges are attached to the daughter with probability $P_n$. The factor $\Omega(k_1,k_2,k_p)$ denotes the probability that $k_1$ edges are inherited among the $k_p$ nodes connected to the parent and $k_2$ edges connect to nodes not connected to the parent,
\begin{equation}
\Omega(k_1,k_2,k_p) =\left(\begin{array}{c} k_p\\ k_1\end{array}\right)
P_e^{k_1}(1{-}P_e)^{k_p{-}k_1} \left(\begin{array}{c} N{-}2{-}k_p\\ {k_2}\end{array}\right)P_n^{k_2}(1{-}P_n)^{N{-}2{-}k_p{-}k_2}.
\end{equation}
Next we consider the effect of the duplication on the adjacent nodes and we need to distinguish between nodes sharing an edge with the parent ($Ed$) and nodes not connect to the parent ($nEd$). Let us first consider the $Ed$ nodes. We introduced above $P_{Ed}(k_p,k,q_E)$ as the probability that a mode, here the parent, of degree $k_p$ is connected to $q_E$ nodes of degree $k$. The duplication process will, with probability $P_e$, attach a new edge from the daughter to each of these nodes and thereby increase their degree from $k$ to $k+1$. Let us now turn to the $nEd$ nodes. There are $D-2-k_p$ nodes which share no edge with the parent. With probability $P_nP_{nEd}(D-2-k_p,k,q_{nE})$ a total of $q_{nE}$ of these nodes are of degree $k$ and will receive a new edged to the daughter. Here $P_{nEd}(D-2-k_p,k,q_{nE})$ is equivalent to $P_{Ed}(D,k,q)$ introduced in Eq. (\ref{P_Ed}), though $P_{nEd}(k_p,k,q_{nE})$ is concerned with the $D-2-k_p$ nodes that a node of degree $k_p$ (in a set of $D-1$ nodes) is {\em not} connected to. Among these $D-2-k_p$ nodes $q_{nE}$ have degree $k$ with probability $P_{nEd}(D-2-k_p,k,q_{nE})$. Therefore we have,
\begin{eqnarray}
\Gamma_{Du}^a(k)&=&
\sum_{q=0}^{D{-}2}\sum_{k_p=0}^{D{-}2}\sum_{\kappa_1=0}^{k_p}
\sum_{\kappa_2=0}^{D{-}2{-}k_p}\sum_{q_1=0}^{\kappa_1}\sum_{q_2=0}^{\kappa_2}
\delta(q_1{+}q_2{-}q)\nonumber \\
&&q\frac{n_{k_p}}{D{-}1}P_{Ed}(k_p,k,\kappa_1)
\left(\begin{array}{c} \kappa_1\\ q_1\end{array}\right)
P_e^{q_1}(1{-}P_e)^{\kappa_1{-}q_1}\nonumber \\
&&P_{nEd}(k_p,k,\kappa_2)
\left(\begin{array}{c} \kappa_2\\  
q_2\end{array}\right)
P_n^{q_2}(1{-}P_n)^{\kappa_2{-}q_2}.\nonumber \\
\label{eq.final}
\end{eqnarray}

\begin{figure}[h]
\centering
\subfigure{
\includegraphics[width=7.5cm]{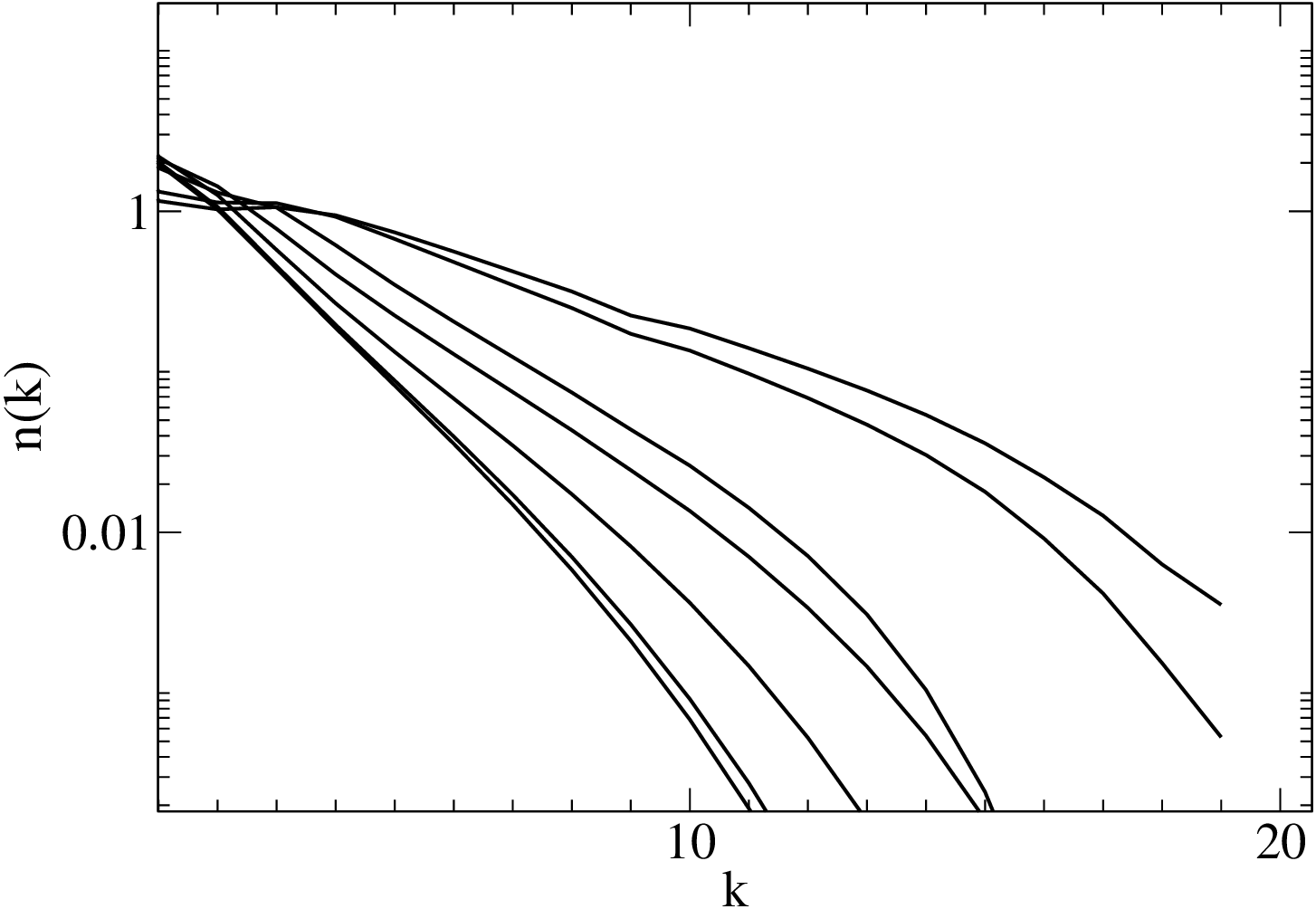}}
\subfigure{
\includegraphics[width=7.5cm]{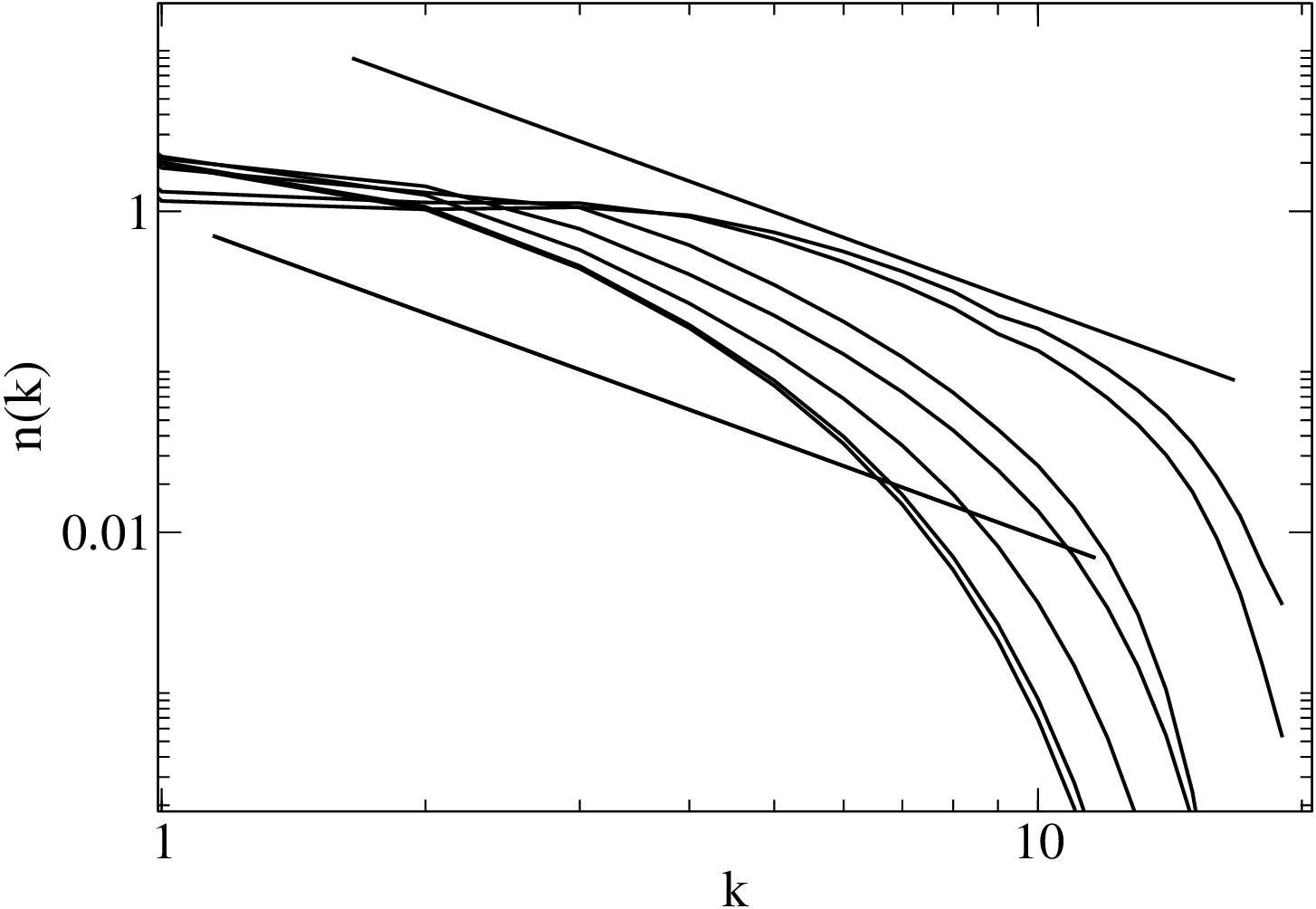}}
\caption{The degree distribution obtained by iteration of the Fokker-Planck equation (\ref{master_D}). The exponential form is visible for a broad range of parameter values in the linear-log plot to the left. The approach towards a $1/k$ dependence in the limit of $P_e\rightarrow 1$ can be seen in the log-log plot to the right. The two straight lines have slope -1. The parameters are $D=20$, $P_e=$ 0.01, 0.1, 0.3, 0.5, 0.7, 0.9, 0.95 and $P_p=$ 0.01. $P_n$ was chosen to be $P_n=P_p(1-P_e)/(1-P_p)$ )}
\label{int_sol}
\end{figure}

Degree-degree correlations induced by the evolutionary dynamics makes is difficult to write an explicit form for  $P_{Ed}(k_1,k_2,q)$ and $P_{nEd}(k_1,k_2,q)$. One can neglect these correlations altogether and try to estimate $P_{Ed}$ and $P_{nEd}$ by purely binomial arguments in the following way. First we deal with $P_{Ed}(k_1,k_2,q)$. The $k_1$ edges emerging from the degree $k_1$ node connects (in this approximation) to nodes of degree $k_2$ with probability $(n_{k_2}-\delta_{k_1,k_2})/(D-1)$ [remember there are $D-1$ nodes when the duplication takes place] hence we use
\beq
P_{Ed}(k_1,k_2,q)=\left(\begin{array}{c}k_1\\q\end{array}\right)\left(\frac{n_{k_2}-\delta_{k_1,k_2}}{D-1}\right)^q\left(1-\frac{n_{k_2}-\delta_{k_1,k_2}}{D-1}\right)^{k_1-q}.
\eeq  
When we treat $P_{nEd}(k_1,k_2,q)$ in the same approximation we obtain $P_{nEd}(k_1,k_2,q)=P_{Ed}(D-2-k_1,k_2,q)$ since we now pick $q$ nodes among the $D-2-k_1$ nodes not connected to the degree $k_1$ node under consideration. One seems, however, to obtain solutions that numerically converge better towards the results obtained by direct simulation (See Fig. \ref{simulations} and Fig. \ref{int_sol}) if one tries to account for the correlations in the allocations of the edges by estimating $P_{ed}(k_1,k_2,q)$ using the following urn argument. We place $M=\sum_kn_k$ edges in an urn. The edges are of two types. Type $\cal A$ edges correspond to the $|{\cal A}|=k_2(n_{k_2}-\delta_{k_1,k_2})$ edges connecting to nodes of degree $k_2$. In addition we have $|{\cal B}|=M-|A|$ type $\cal B$ edges connecting nodes of degree different from $k_2$. The probability that among $k_1$ randomly picked edges $q$ are of type $\cal A$ and $k_1-q$ are of type $\cal B$ is given by
\beq
P_{Ed}(k_1,k_2,q)=  \left(\begin{array}{c} k_1\\ q\end{array}\right)
\big(\frac{k_2n_{k_2}}{M}\big)^q
\big(1-\frac{k_2n_{k_2}}{M}\big)^{k_1-q}.
\eeq
Again we assume $P_{nEd}(k_1,k_2,q)=P_{Ed}(D-2-k_1,k_2,q)$. In general it is not simple to find analytic solutions to this somewhat involved set of equations. The result of iterating the Fokker-Planck Eq. (\ref{master_D}) using these estimates is shown in Fig. \ref{int_sol} for diversity $D=20$, which makes the numerical iteration manageable. We notice good qualitative agreement with the behaviour of simulation results presented Fig. \ref{simulations}. For a broad range of values of the parameters the degree distributions exhibit much of the same approximate exponential form as produced by the individual based Tangled Nature model described in the Sec. \ref{TaNa_model}.
\vspace{1cm}

Let us finally mention that direct simulations \cite{lair06:stea} of the simple model introduced in section \ref{simple_model} show that in the limit $P_e\rightarrow 1$, $P_n\rightarrow 0$ and $P_p\ll 1$ the degree distribution $n_k$ behaves like $1/k$. The Fokker-Planck equation Eq. (\ref{master_D}) confirms this result. In the limit $P_e=1$ and $P_n\rightarrow 0$ (i.e. perfect replication) the Fokker-Planck equation reduces to
\bea
n_k(t+1)&=&n_k(t)+n_k(\frac{1}{D-1}-\frac{1}{D})+\frac{2P_p}{D-1}(n_{k-1}-n_k)\nonumber\\
&&+\sum_{k_1=1}^{D-2}\sum_{q=1}^{k_1}qn_{k1}[\frac{1}{D}P_{Ed}(k_1,k+1,q)-\frac{1}{D-1}P_{Ed}(k_1,k,q)\nonumber\\
&&-\frac{1}{D}P_{Ed}(k_1,k,q)+\frac{1}{D-1}P_{Ed}(k_1,k-1,q)].
\eea
Including only the leading terms from $k_1=1$ and $q=1$ one obtains
\bea
n_k(t+1)&=&n_k+\frac{n_k}{D(D-1)}+\frac{2P_p}{D-1}[n_{k-1}-n_k]\nonumber\\
&&+\frac{n_1}{M}[\frac{1}{D}\{(k+1)n_{k+1}-kn_k\}-\frac{1}{D-1}\{kn_k-(k-1)n_{k-1}\}].
\eea
In the limit $D\gg1$ and $P_p\ll1$ this equation has the stationary solution $n_k\propto 1/k$.

\section{Summary and Discussion}
We have considered networks, consisting of a constant number of nodes, with the allocation of edges evolving according to dynamical rules inspired by biological reproduction. We discussed mathematically the mechanisms behind the exponential form of the degree distribution found typically except in the limit where the reproduction doesn't involve any mutations at all. The Fokker-Planck equation we derived allows us to verify that the degree distribution changes from exponential to a power law in the limit where the replication is of high fidelity. 

The generality of the dynamics may make the results relevant to a range of problems including protein-protein interacting networks and routers dynamics in communication networks.

\section{Acknowledgement}
I am grateful for very fruitful collaboration with with Dr Simon Laird.

\end{document}